# Uncommon clustering in dilute Ti-Fe alloys

D. W. Boukhvalov[1,2,*], Yu. N. Gornostyrev,[2,3,4] and M. I. Katsnelson[5,2]

[1]*College of Science, Institute of Materials Physics and Chemistry, Nanjing Forestry University, Nanjing 210037, PR China*
[2]*Theoretical Physics and Applied Mathematics Department, Ural Federal University, Mira Street 19, 620002 Ekaterinburg, Russia*
[3]*Institute of Metal Physics, UB of RAS, S. Kovalevskaya st.,18, Ekaterinburg 620990, Russia*
[4]*Institute of Quantum Materials Science, Ural Hi-Tech Park, Konstruktorov st, 5, Ekaterinburg 620072, Russia*
[5]*Radboud University, Institute for Molecules and Materials, Heyendaalseweg 135, Nijmegen, 6525 AJ, Netherlands*

We present the results of ab initio modeling of structure of dilute Ti-Fe, a typical representative of quenched Ti-based transition-metal alloys. We have demonstrated that beyond the solubility limit this alloy cannot be described in common terms of substitutional and interstitial alloys. Instead, very stable local clusters are formed in both low-temperature hcp and high-temperature bcc phases of alloys, with almost identical local structures. This gives an example of geometrically frustrated state and explains unusual concentration behavior of Mössbauer spectra discovered long ago for this system.

E-mail: danil@njfu.edu.cn

## 1. Introduction

Titanium-based alloys of transition metals have many important applications [1]. They demonstrate unusual (for metallic systems) properties such as negative temperature coefficient of resistivity, pseudogap in infrared optical spectra, strong concentration anomalies of sound velocities and attenuation, etc. [2-7]. Titanium is a polymorphic metal with low-temperature hcp α-phase and high-temperature bcc β-phase (under pressure, also ω-phase [8] arises). Transition metals to the right of Ti in the Periodic Table are β-stabilizers. At high enough dopant concentration, a special structural state appears in quenched alloys with coexistence of β- and athermal ω-phase [1,8]. Studies of Mössbauer spectra of $Ti_{1-x}Fe_x$ alloys [9] gave an unexpected result, namely, insensitivity of local surrounding of iron impurity to the global crystal order. Both quadruple splitting and isomer shift characterizing local structure and electronic structure, respectively, turn out to be almost identical in hcp α-phase ($x<0.040$) and bcc β-phase ($x>0.070$) and almost concentration-independent within these phases. It was suggested in Ref.[10] based on these experiments that quenched $Ti_{1-x}Fe_x$ alloys are neither substitutional nor interstitial and that already in high-temperature β-phase Fe forms, together with surrounding Ti, quite stable local clusters which are almost not reconstructed under the quenching to different low-temperature phases. The supposed cause was a geometric frustration: interatomic distances in both α- and β-Ti are much larger than in the most stable intermetallic phase TiFe with CsCl structure and therefore just substitution of Ti by Fe is not optimal energetically. The assumption [10] is quite unusual for crystalline metallic alloys. Electronic structure calculations show that Fe as a substitutional impurity should be magnetic [11-13], which seems to agree with experiment [12] for a very small concentration of Fe, before the solubility limit $x_c \approx 10^{-3}$ @RT, but clearly contradicts experimental data for $x > x_c$ [3]. Only interstitial position of Fe was considered as an alternative in electronic structure calculations up to now [12,13]. Here we present detailed first-principle calculations which confirm the hypothesis [9,10] on clustering of Fe in quenched Ti-Fe alloys with a local structure of the clusters that is different from those dictated by crystal lattice structure of the matrix. This means that even for so chemically simple and non-exotic alloys as Ti-Fe the conventional separation of metallic alloys into two classes, substitutional and interstitial, is not enough and much more complicated local structural state can be formed. More specifically, this state can be described as a formation of complexes of more or less substitutional Fe impurities with interstitial Ti atoms.

## 2. Methods

The modeling was performed by density functional theory (DFT) in the pseudopotential code SIESTA [14], as was done in our previous work on a similar subject [15]. All calculations were performed using the generalized gradient approximation (GGA-PBE) with spin-polarization [16]. The ion cores were described by norm-conserving pseudo-potentials [17] and the wave functions are expanded with a double-$\zeta$ plus polarization basis of localized orbitals for iron and titanium. Full optimization of the atomic positions was performed, and the forces and the total energy were calculated with the accuracy of 0.04 eV/Å and 1 meV, respectively. For the modeling of all configurations the 3×3×3 supercell of 54 titanium atoms in hcp and bcc structures was used. To check the effect of supercell size, larger supercells with 96 titanium atoms in hcp and 128 atoms in bcc configuration were used. All the calculations were carried out with an energy mesh cut-off of 300 Ry and a k-point mesh of 6×6×4 (3×3×2 for larger supercell) and 6×6×6 (3×3×3 for larger supercell) in the Monkhorst-Pack scheme [17] for α and β phases, respectively. For the plots of DOS the k-point mesh was increased up to 8×8×6 and 8×8×8, respectively. Formation energy of considered configurations was calculated by the standard formula: $E_{form} = E_{nFe+mTi} - (nE_{Fe} + mE_{Ti})$, where $E_{nFe+mTi}$ is the total energy of the supercell contain m atoms of Ti and n atoms of Fe, $E_{Fe}$ – total energy per atom of iron in α-Fe, and $E_{Ti}$ is the total energy per titanium atom in the corresponding phase (α or β).

## 3. Computational results

The first step of our calculations is the check of energetically preferable positions of iron impurities in α–titanium at various concentrations. We examine four possible configurations: (i) quasi–random distribution of substitutional iron impurities, (ii) aggregation of substitutional iron impurities with further formation of clusters of substitute atoms (nFe(S)) where n is the number of iron atoms in supercell), (iii) interstitial iron impurities with further formation of cluster of substitutional iron impurities around interstitial one (nFe(S)+Fe(I)), (iv) formation of clusters of substitutional iron impurities around atom of titanium in interstitial void (nFe(S) + Ti(I), see Fig. 1a,b). We examined all these configurations; further we discuss the results only for the structures with the lowest formation energies.

### 3.1. α-phase

The computational results (Fig. 2) allow us to suggest energetically optimal configurations of iron impurities dependent on their concentration. First of all, iron is more soluble in β-Ti than in α-Ti, in agreement with the previous computations [11] and the well-known fact that Fe is β-stabilizer [1]. In α-Ti at the lowest studied concentrations ($x<0.06$) formation of single substitutional impurity (~0.48 eV/Fe) and pairs of substitutional iron atoms is more energetically favorable than other types of defects (curves 2,3 in Fig. 2a). Formation of the single and double defects does not provide visible changes in lattice parameters of the system. For concentration of Fe in α-phase roughly between 4 and 10 at% a quasi–random distribution of substitutional impurities is not energetically favorable; note however that α-β transition happens at the average concentration Fe near 4at% at room temperature [1,5,9]. Thus, we can consider the concentration of 4 at% as the lowest margin of dilution. In the samples of α-Ti with concentrations of Fe impurities 6-12 at% the most energetically favorable configuration is the cluster of substitutional impurities around interstitial titanium atom which formation energies are of the order of 0.1~0.2 eV/Fe atom lower than for other configurations (see Fig. 2a). Note that the formation energy of interstitial Ti-atom without substitutional Fe-impurity in its vicinity is rather high (~2.1 eV/Ti), thus, the formation of Fe(S)-Ti(I) bond decreases the energy of this defect. To check the effect of supercell size we performed also the calculations for larger supercell of 96 atoms for the different configurations of 6 Fe-impurities and found that the difference between the formation energies configurations remains almost the same (deviations within 0.02 eV). Formation of the clusters in the supercell of 54 atoms provides a decrease of the lattice parameter $a$ at 1% and extension of the cell size along $c$ axis about 6%. For the supercell of 96 atoms these values decrease to 0.3 and 2.4%, respectively.

To clarify the nature of these clusters (see Fig. 1b) we have checked interatomic distances and have found that for the clusters nFe(s)+Ti(I) the Fe–Ti distance almost coincide with the equilibrium values for intermetallic TiFe (B2 structure), 2.59~2.61 and 2.58 Å, respectively, in contrast to other types of impurity distributions at intermediate concentrations where deviation from the values for TiFe is more than 0.15 Å. The values of Fe-Ti-Fe angles in nFe(s)+Ti(I) clusters is about 65.6~66.2 ° that is close to the corresponding value in B2-TiFe crystals. This confirms a suggestion [10] that geometric factors play a crucial role in the structure of Ti-based alloys above the solubility limit. Thus, we can describe nFe(s)+Ti(I) clusters as the smallest nuclei of distorted TiFe phase. At equilibrium, the system is a two-phase, with coexisting TiFe and α-Ti phases [1]. In the quenched state, there is nanoscale inhomogeneity of the type described above. In general, this inhomogeneous state can be stabilized by misfit strains [19]. Microscopic description of these phenomena requires calculations for specific materials, and this is what we have done for Ti-Fe.

Noticeable experimental result is the absence of local magnetic moments on Fe in the concentration range under consideration [3,12,13]. From all four studied models, only nFe(s)+Ti(I) clusters have zero magnetic moments, other three studied configurations provide an appearance of magnetic moments on iron impurities of the order of 2.2 $\mu_B$/Fe. The other experimental fact is the formation of pseudo–gap in electronic structure in diluted Fe–doped titanium [2-6]. The calculated electronic structures for nFe(s)+Ti(I) clusters (Fig. 3a) do demonstrate similarity of the electronic structure of iron and Ti(I) with electronic structure of B2-FeTi (Fig. 3c). Note that the difference in electronic structure of Ti atom inside the iron cluster is rather different from the electronic structure of other Ti-atoms in the system. Due to rather low concentration of the impurities, the electronic structure of the whole system (total DOS) is mainly determined by the contribution from titanium atoms (Fig. 3a,b). The formation of the clusters decreases the density of states at the Fermi level, which can be related to the experimentally observed pseudo-gap formation in these systems [2,4-6]. Therefore, we can say that all structural, magnetic and electronic properties of nFe(s)+Ti(I) clusters are in a qualitative agreement with the experimental results.

### 3.2. β-phase

Mössbauer spectra [9] demonstrate that local Fe–Ti structures with similar characteristics exist in both α- and β-Ti phases, with quadrupolar splitting and isomer shift being almost concentration independent (except a close vicinity of the structural transition point [9]). Importantly, both these phases, α and β, at room temperature and below are obtained by quenching of the high-temperature phase (β-Ti).

In contrast to hcp α–Ti, in bcc β–Ti the interstitial voids are too small to allow appearance interstitial Fe impurities. Therefore, we start out modeling from the pair of impurities and consider only three types of configurations: (i) quasi–random distribution of iron atoms, (ii) clusters of substitutional iron impurities, and (iii) aggregation of iron impurities in the vicinity of one titanium atom (Figs. 1c,d) which further could be transform to nFe(s)+Ti(I) clusters in α–Ti (see below). Formation energy of the single impurity in β-Ti is -1.96 eV, that is, corresponds to instability of the host system. Incorporation of the single iron impurity does not change lattice parameters of the supercell. At the same time, at $x = 0.04 - 0.05$ quasi-random distribution of impurities turns out to be the most favorable (see Fig. 2b). Note that experimentally [9] in this region, contrary to large and smaller Fe concentration, the Mössbauer spectra cannot be described as a single, quadruple-split doublet and demonstrate a broad distribution of observable Fe positions. In this respect, our results also seem to agree with the experiment. For the case of 4 at% Fe (two iron impurities per 54 atomic supercell) the formation of Fe-Ti-Fe clusters is about 0.4 eV/Fe less energetically favorable and formation of Fe-Fe neighbor's pairs is 0.8 eV less preferable than the random configuration. We do not show the results for $x < 0.04$ since bcc phase does not exist at these concentrations. Formation of Fe-Ti-Fe clusters of six iron atoms provides unilateral expansion of the supercell at 3.2%. To check the effect of supercell size we performed also the calculations for larger supercell of 128 atoms for the different configurations of 6 Fe-impurities and found that the difference between formation energies of the configurations under consideration remain almost the same (deviations within 0.02 eV).

These results indicate a non-monotonous dependence of the interaction energy on the distance between iron impurities in β-Ti. To explore this issue in detail we have calculated the interaction energies between two iron impurities at the distance $X_i$ corresponding to $i$-th coordination shell, $E_{int}(X_i) = E_{tot}((N-2)Ti,2Fe,X_i)-2E_s$. Here $E_{tot}((N-2)Ti,2Fe,X_i)$ is the energy of the crystallite where two Ti atoms are substituted by Fe, $E_s$ is the solution energy of a single Fe atom in bcc Ti. We found atypical non-monotonous behavior of the interaction energy $E_{int}(X_i)$; Fe-Fe attraction increases with the distance, reaching the highest absolute values for $4^{th}$ coordination spheres with a rapid further decreasing of the magnitude (Fig. 4). This behavior indicates the competition of two contributions to the interaction energy of Fe impurities $E_{int}(X_i)$ related to the chemical bonding and lattice distortions.

Due to these features of Fe-Fe interaction the increase of impurity concentration from 4 to 15% makes the formation of iron clusters around titanium atom (Fig. 1d) the most energetically favorable. For higher concentration of iron the random distribution of impurities makes appearance of Fe-Fe pairs unavoidable and aggregation of impurities around one Ti atom turns out to be the best choice to prevent the presence of unfavorable Fe-Fe pairs. This explains peculiarities of the Mössbauer spectra near the transition point, with a broad distribution of observable iron positions [8]. For higher (in α-phase) and lower (in β-phase) concentrations there is an unique favorable atomic configuration. Now we will describe this configuration in more detail.

Fe–Ti distances in the energetically favorable clusters in β-Ti are in the range of 2.59~2.61 Å and angles close to 70°, like in α-phase (see Fig. 1b,d), that suggests a tendency to formation of very stable local clusters of distorted TiFe phase in both hosts. It turns out that only these clusters in β-Ti among all studied ones are nonmagnetic (other considered configurations provide appearance of magnetic moments about 2.4 $\mu_B$/Fe). Electronic structure of clusters with iron surrounding Ti atom (Fig. 3b) is similar to the electronic structure of nFe(s)+Ti(I) clusters in α-phase and electronic structure of the Ti-atoms inside the cluster is similar to that in B2-FeTi. There is a clear tendency to the formation of pseudo–gap in the density of states near the Fermi energy.

### 4. Discussion and conclusions

Based on the presented results we can speculate about possible mechanism of survival of Fe clusters during the bcc-hcp transformation when cooling from high temperatures. As commonly accepted, the bcc-hcp transformation in Ti during the cooling is realized by phonon mechanism ensuring the reconstruction of bcc lattice under the Burgers scheme [20]. This scheme involves two types of distortions: (i) opposite displacement of adjacent $(110)_{bcc}$ planes in the $[110]_{bcc}$ direction and (ii) volume-conserving shear deformation in the $[001]_{bcc}$ direction, keeping the distance between the $(110)_{bcc}$ planes unchanged. In the case of the clusters with robust and rigid Fe-Ti bonds, the titanium atom in the center of cluster will be inherited upon the transformation, wherein the center of the cluster remains in initial position corresponding to the formation of interstitial defect in hcp lattice. Our computational results for the intermediate steps along the Burgers path for the case of cluster of six iron atoms around Ti center (Fig. 5) demonstrate that the values of Fe-Ti-Fe angles change smoothly during the hcp-bcc transition and the values of Fe-Ti distances slightly (less than 0.1 Å) increase at the intermediate stages of the transformation. It confirms the structural stability of discussed clusters during the transformation of bcc to hcp Ti, wherein the 6Fe(s)+Ti(S) clusters are rearranged in to 6Fe(s)+Ti(I) ones.

Based on the results of our modeling one can conclude that quenched Ti-Fe alloys above the solubility limit can be considered for $0.06 < x < 0.13$ as neither substitutional nor interstitial alloys. Instead, a formation of local clusters with close to optimal Ti-Fe interatomic distances takes place. This explains mysterious concentration evolution of the Mössbauer spectra [9] of these alloys and supports a hypothesis [10] on the formation of locally symmetry broken structure. At the same time, we specify this hypothesis and suggest a model for real structural state of these very common and practically important alloys.

MIK acknowledges a financial support by NWO via Spinoza Prize. The work was supported by Act 211 Government of the Russian Federation, contract No. 02.A03.21.0006. DWB acknowledge support from the Ministry of Science and Higher Education of the Russian Federation, Project № 3.7372.2017/8.9

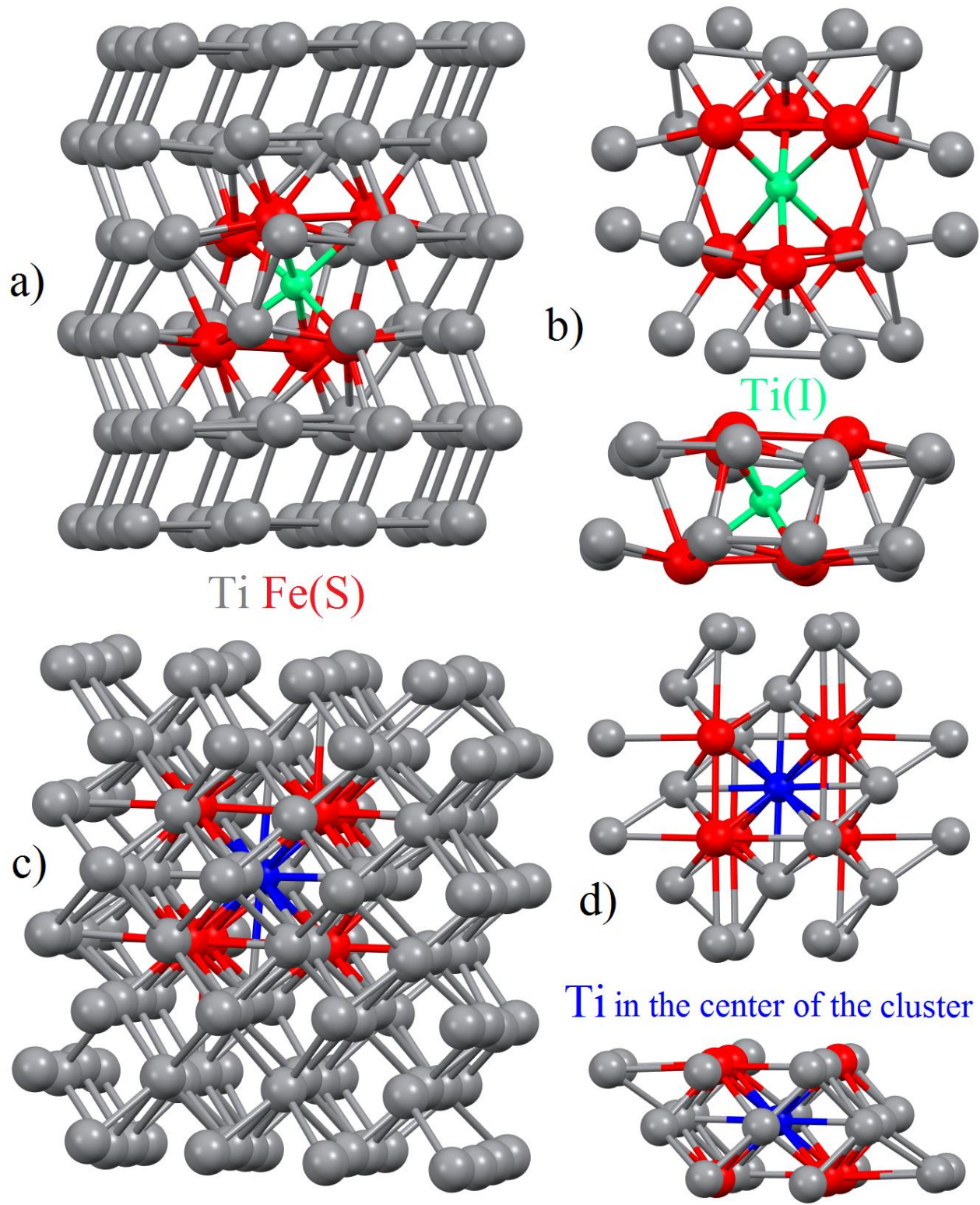

**Figure 1** (a), (b): Optimized atomic structures of 96 atoms supercell of α-Ti with 6Fe(S)+Ti(I) cluster. Panel (b) shows the environment of the same cluster but from some different perspectives. (c), (d): Optimized atomic structure of 128 atoms supercell of β-Ti with six Fe(S) impurities clustered around titanium atom. Panel (b) shows the environment of the same cluster but from some different perspectives.

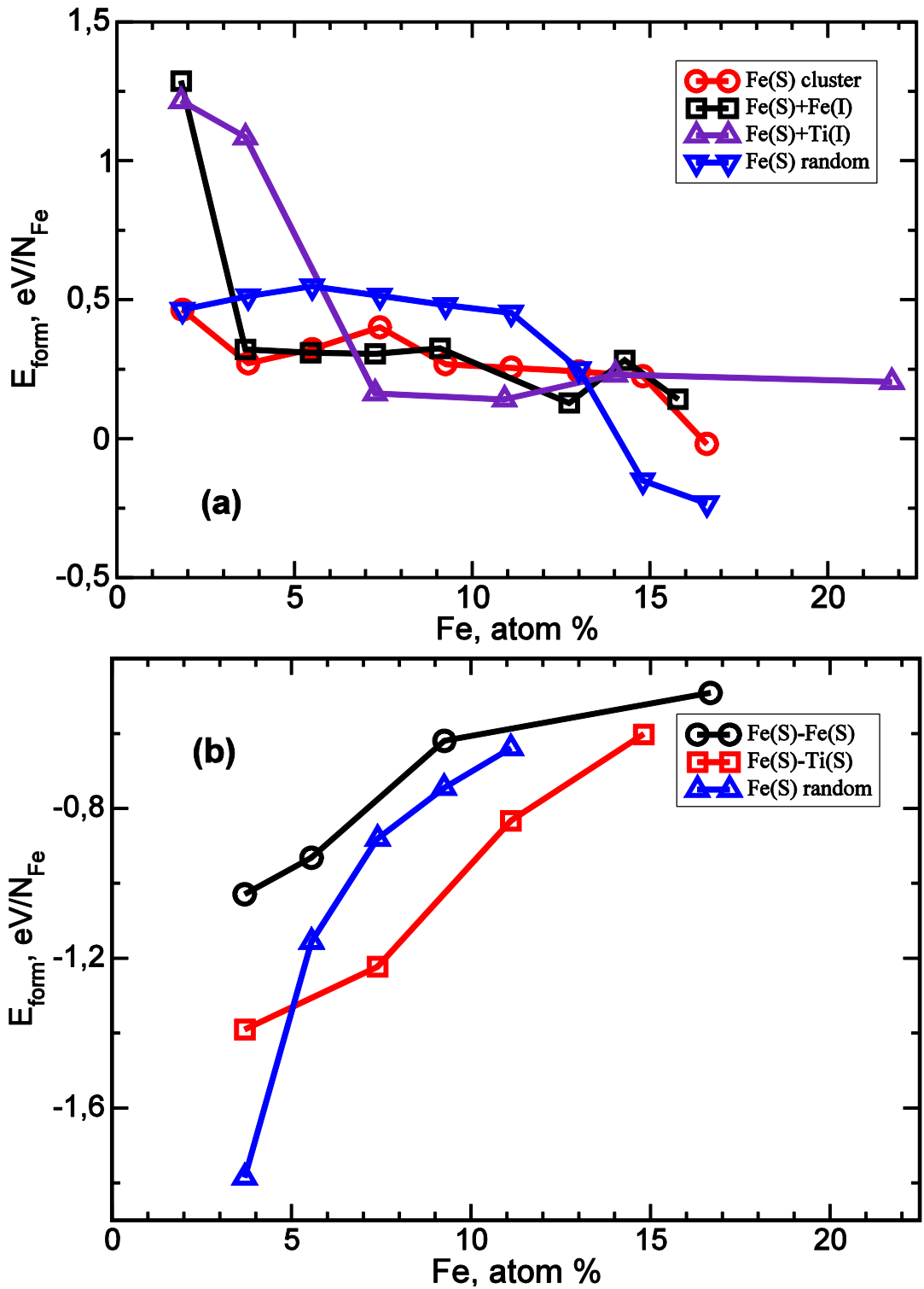

**Figure 2** Formation energies of various configurations of impurities (see the text) in α (a) and β (b) phases of titanium.

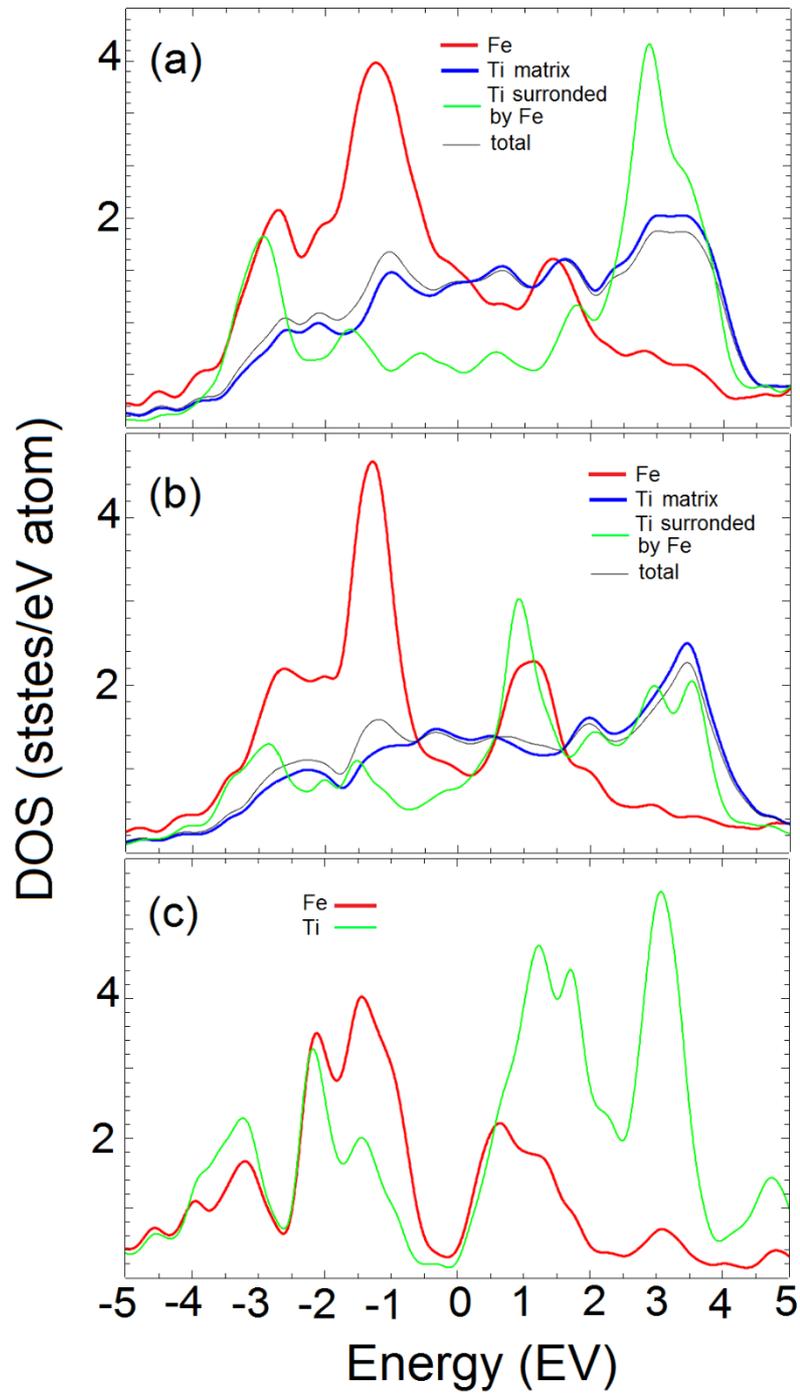

**Figure 3** Total densities of states per atom and partial densities of states per atom of iron and titanium, and for titanium atom surrounding by iron atoms for the most energetically favorable configurations of impurities at given concentration about 11% (six substitutional iron impurities around titanium atom) in α (a) and β (b) phases of titanium and (c) B2-FeTi.

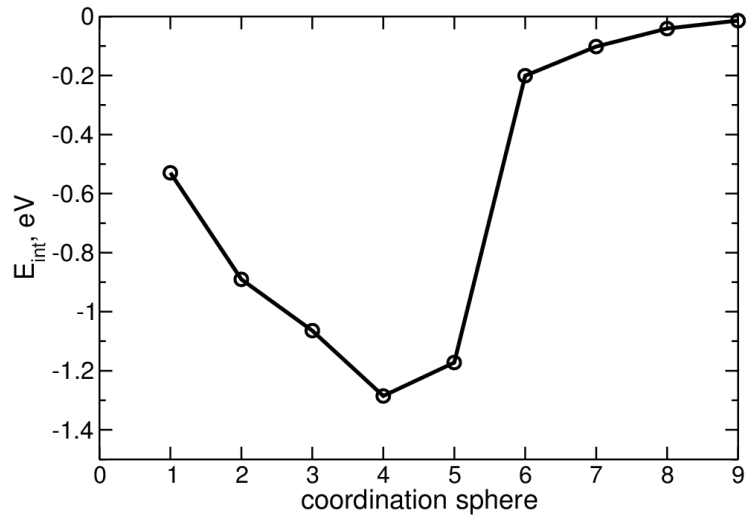

**Figure 4** Interaction energy of the pair of iron impurities in β-Ti as function of coordination sphere of disposition of second defect in the pair.

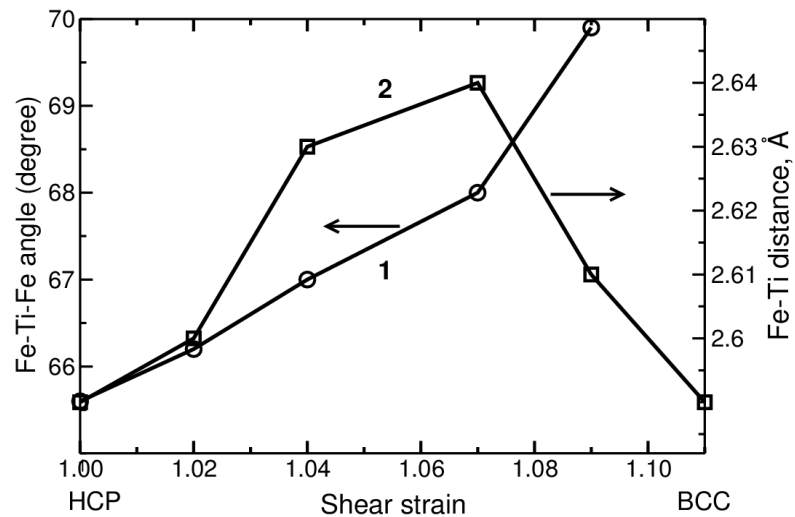

**Figure 5** Changes of interatomic angles (1) and distances (2) in cluster of six iron atoms around titanium center as function of lattice distortion of during hcp-bcc transition along the Burgers path [18].